\documentclass[12pt,amsmath,amssymb,floatfix]{revtex4}
\usepackage{color}
\input{epsf} 
\def\beq{\begin{equation}} 
\def\eeq{\end{equation}} 

\begin{document}

\begin{titlepage}
\renewcommand{\thefootnote}{\fnsymbol{footnote}}
\begin{flushright}
\end{flushright}

\begin{center}
\Large{
{\bf Next-to-leading order QCD corrections to\\} 
\vspace*{0.3cm}
{\bf hadron+jet production in $pp$ collisions at RHIC\\}}
\vskip 1.cm
{\large  Daniel de Florian }  \\
\vspace*{0.3cm}
\normalsize
 Departamento de F\'\i sica\\  
                       Facultad de Ciencias Exactas y Naturales \\
                       Universidad de Buenos Aires \\
                       Pabell\'on I, Ciudad Universitaria \\ 
                       (1428) Capital Federal \\
                       Argentina
\end{center}
\vskip 1.cm
\begin{center}
{\bf Abstract} \\
\end{center}
\vspace{0.2cm}
\normalsize
We compute the next-to-leading order (NLO) QCD corrections to the 
spin-independent and spin-dependent cross sections for the production of a single-hadron
accompanied by an opposite jet in hadronic collisions.
 This process is being studied experimentally at RHIC, providing a
 new tool to unveil the polarized gluon distribution $\Delta g$.
 We perform a detailed analysis of the phenomenological
impact of the observable at NLO accuracy and show that the preliminary data
by the STAR collaboration confirms
the idea of a small gluon polarization in the  $0.05\lesssim x\lesssim 0.3$ range.
April 2009 
\end{titlepage}
\section{Introduction}

In the past two decades, measurements of the spin asymmetries 
$A_1^N$ ($N=p,n,d$) in longitudinally polarized deep-inelastic scattering 
(DIS) \cite{reviewexp} have provided new information on the spin structure of the 
nucleon. One of the most surprising results is that only a small fraction of the
spin of the proton can be attributed to the spin of the quarks.
The main goal of the spin program, besides obtaining the partonic share to the total 
spin of the nucleon, 
 is the extraction of  
 the full set of the $x-$dependent polarized quark ($\Delta q = q^{\uparrow}-q^{\downarrow}$) 
and gluon ($\Delta g=g^{\uparrow}-g^{\downarrow}$) densities of the nucleon. 
Many phenomenological analyses \cite{reviewth} 
demonstrate, however, that available DIS data alone is not sufficient for this purpose.
This is true in particular for $\Delta g(x,Q^2)$ since it contributes to 
DIS in leading-order (LO) only via the $Q^2$ dependence of $g_1$ (or $A_1$) 
which could not yet be 
accurately studied experimentally. As a result of this, it turns out 
 that the $x$ shape  of $\Delta g$ seems to be hardly 
constrained by the DIS data.

The precise extraction of $\Delta g$ thus remains one of the most interesting 
challenges for spin physics experiments. 
The RHIC collider at BNL, running in 
a  proton--proton mode with longitudinally polarized beams provides
the ideal tool for that purpose. The observables measured so far include 
single-pion production at center-of-mass energy $\sqrt{s}=62$ \cite{Adare:2008qb} and 200 GeV \cite{Adare:2007dg} 
and jet production at  $\sqrt{s}=200$ GeV \cite{Abelev:2007vt}. Unlike DIS, those processes
have a direct gluonic contribution already at the lowest order.

A next-to-leading order (NLO) global analysis that includes all available data 
from inclusive and semi-inclusive polarized deep-inelastic scattering, as well as 
from polarized proton-proton scattering at RHIC
has been recently performed \cite{dssv}. The main outcome of the analysis is the indication of
a rather small gluon polarization in the nucleon over the limited region of momentum fraction
$0.05\lesssim x \lesssim 0.2$. 

Recently the STAR collaboration at RHIC has presented preliminary data \cite{adam} corresponding to
the 2006 run on a less
inclusive observable, involving the production of a charged hadron accompanied by a back-to-back jet. 
From the pure experimental point of view, counting with an opposite jet allows one to use it as a trigger 
for the hadron, reducing the bias in the selection. Furthermore, having a more exclusive observable,
and particularly counting with the transverse momentum of both the hadron and the jet, permits  one to
perform a more detailed study to extract $\Delta g$.

In order to make reliable quantitative predictions for a high-energy process,
it is crucial to determine the NLO QCD corrections
to the Born approximation. In general, in hadronic collisions, cross sections computed 
at the lowest order in perturbation theory are severally affected by the
dependence on the `unphysical' factorization and renormalization scales, dependence that can be 
partially cured only by  including the NLO corrections.
Furthermore, the appearance of one extra final-state parton in the NLO from the 
$2\rightarrow 3$ real corrections allows one to improve the 
matching between the theoretical calculation and the realistic 
experimental conditions, particularly when jets are present.

The calculation of the NLO QCD corrections to hadron+jet production by unpolarized and polarized
hadrons is the purpose of this paper.  Several
modern versions of the subtraction method to calculate any infrared-safe quantity in 
unpolarized collisions are at present available in the  
literature~\cite{CS,FKS}. The formalism of Ref.~\cite{FKS}
has been used in Ref.~\cite{Jets97} to construct a Monte Carlo code
that can calculate any jet infrared-safe observable
in hadron--hadron unpolarized collisions and generalized,  in Refs.~\cite{jetpol,jetpol-photo}, to the polarized case.
In \cite{singlehad}, the method was extended to the case of singe-hadron inclusive observables.

 In the present paper, we 
apply the method of Refs.~\cite{FKS,Jets97,singlehad}  to the case of 
hadron+jet observables.
 As a result, we will 
present a customized code, with which it will be possible to calculate 
any infrared-safe quantity corresponding to one-hadron+jet production 
 to NLO accuracy, for both polarized and unpolarized collisions.
 With the technique introduced in \cite{dssv} the code is suited 
 to allow the inclusion of the observable in a global fit analysis.
 
With the code at hand, we analyze in detail the phenomenological implications of the observable.
The key point here is that, being a more exclusive measurement, it is possible to analyze the data
in terms of a new set of variables and, also, to
impose experimental cuts that can enhance the contribution of some partonic subprocesses over others.
That is fundamental in order to increase the sensitivity on the spin-dependent gluon distribution
in polarized collisions.

This paper is organized as follows: in section 2 we describe the main ingredients of the calculation
and study the perturbative stability of the NLO results, by looking at the scale dependence 
and `$K$-factors'. 
In section 3 we discuss some phenomenological aspects of hadron+jet production in hadronic
collisions and in section 4 we present the results for the asymmetries at RHIC .
Finally, section 5 contains the conclusions.

\section{NLO corrections and Perturbative stability}  
\setcounter{equation}{0}
  
The factorization  theorem~\cite{CSS} allows one to write the cross section for 
one-hadron production in hadronic collisions as
\begin{eqnarray}
\hspace{-2cm} 
\label{eq:fact}
{d \sigma^{pp\rightarrow h+jet X}} &=& \hspace*{-0.2cm}
\sum_{f_1,f_2,f} \int \hspace*{-0.1cm} 
dx_1\, dx_2\, dz \,\,{f_1^{H_1}} (x_1,\mu_{FI}^2)\,\, {f_2^{H_2}} (x_2,\mu_{FI}^2)  \nonumber \\ &&
\times\, {d{\hat{\sigma}}^{f_1 f_2\rightarrow fX'}} 
(x_1\, p_1,x_2\,p_2,p_{h}/z,\mu_{FI},\mu_{FF},\mu_R)\,\, {D_f^{h}} (z,\mu_{FF}^2) \times\, {\cal S}(p_i)
\end{eqnarray}
where, $H_1$ and $H_2$ are the colliding particles with momentum $p_1$ and $p_2$, respectively, 
$h$ is the outgoing hadron with momentum $p_h$ and the sum in Eq.(\ref{eq:fact}) runs over all 
possible  initial and final partonic states.
The parton distributions $f_i^{H_i}$ are evaluated at the factorization scale $\mu_{FI}$, the
fragmentation functions at the scale $\mu_{FF}$ \footnote{We chose equal factorization 
scales $\mu_{FF}\equiv\mu_{FI}\equiv\mu_{F}$} and the coupling constant, appearing in the 
perturbative expansion of the partonic cross section, at the renormalization scale $\mu_{R}$.
The measurement function ${\cal S}(p_i)$ accounts for possible experimental cuts applied to 
the cross section, and in this particular case,
for the definition of the jet in terms of the kinematics of the final-state partons. 
The analogous of  Eq.(\ref{eq:fact}) for polarized cross section is obtained  by replacing the parton 
distributions and the partonic cross section by its polarized expressions, $\Delta f_i^{H_i}$ and 
${d{\Delta \hat{\sigma}}^{f_1 f_2\rightarrow fX'}}$, respectively.
As usual, the (longitudinally polarized) asymmetry is defined by the ratio between the polarized and 
unpolarized cross sections
\begin{eqnarray}
{A}_{LL}^{h}=\frac{d\Delta\sigma}{d\sigma} \, .
\label{asypt}
\end{eqnarray}
In order to evaluate the QCD corrections to the process we rely on the version of the subtraction 
method introduced and extensively discussed in Refs.\cite{FKS,Jets97}, 
and in Ref.~\cite{jetpol}. We refer the reader to those references for the details.
The implementation is performed in a MonteCarlo like code, profiting from the one  
available for the computation of single-hadron production in  Ref.\cite{singlehad}.
Since the calculation in \cite{singlehad} provides the 
full kinematics for the final-state partons it is possible, after some modifications, to build the jet
kinematics in terms of them and obtain the required cross section.
It is worth noticing that the same code computes both unpolarized and polarized cross sections, 
since the formal structure of the corrections is exactly the same.

In this work, we concentrate on  the phenomenology of pion production accompanied 
by a back-to-back jet
for the kinematics of the STAR experiment at RHIC with a center-of-mass energy of $\sqrt{S}=200$ 
GeV. Unless otherwise stated, we require the pion  transverse momenta to be larger than $2$ GeV 
and the one for the jet to $10\, {\rm GeV} < p_T^{jet} < 25$ GeV. 
The rapidities of the pion and the jets are limited to the range $|\eta|<1$. They have to be 
separated in the azimuthal angle by $\Delta \phi\equiv |\phi^{\pi}-\phi^{jet}| > 2$ to ensure 
the pion and jets are produced from `opposite-side' partons. Finally, the jets are defined according to 
the cone algorithm with R=0.7.
\begin{figure}[htb]
\vspace{0.5cm}
\begin{center}
\begin{tabular}{c}
\epsfxsize=8truecm
\epsffile{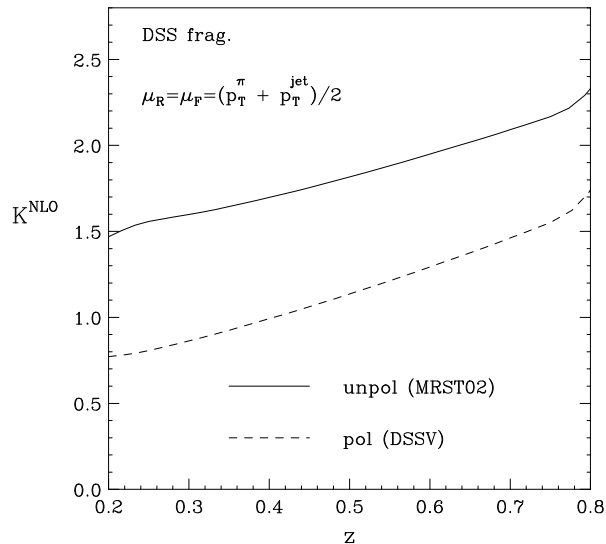}\\
\end{tabular}
\end{center}
\vspace{-0.5cm}
\caption{{\em \label{k} Unpolarized (solid) and polarized (dashes) NLO K factors. The choice
of the  factorization and renormalization scales corresponds to
$\mu_F=\mu_R=\left( p_T^{\pi} +p_T^{jet}\right)/2$. }}
\end{figure}

The size of radiative QCD corrections to a given hadronic
process is often displayed in terms of a $K$-factor which represents the
ratio of the NLO over LO results. In the calculation of the numerator of
$K$, one obviously has to use NLO-evolved parton densities. As far as the
denominator is concerned, a natural definition requires the use of
LO-evolved parton densities. 
In the polarized case, a problem arises for such a definition: since the polarized pdfs are not as
well constrained as the unpolarized ones, it might happen that quite different results for the
 $\Delta f$'s can emerge when the fit is performed at LO or at NLO. This is particularly enhanced
 by the fact that the polarized pdfs have nodes.
 Therefore,
the $K$-factor for a given process, defined using LO parton densities in the 
denominator could be largely affected by the fact that some polarized  densities are at 
present not well constrained.
In order to avoid this problem, we define 
the $K$-factor as the ratio between the NLO and the `Born' cross section, where the 
latest corresponds to the use of NLO-evolved parton densities (and two-loop expression 
for $\alpha_s$) when evaluating the 
 lowest-order partonic cross sections in the denominator.
Nevertheless, it is important to remember that the $K$-factor is {\it not} a physical 
quantity and just provides a number to 
`quantify' the effect of the higher-order corrections.

In the unpolarized case, we use the MRST2002 parton distributions 
\cite{mrst2002}. Differences of the order of percent in the cross section are observed when
 more recent distributions are considered. Nevertheless,
since the DSSV \cite{dssv} distribution we use in the polarized case set is obtained from a global analysis
that relies on the unpolarized MRST2002 set as a reference, we will restrict the analysis 
to the MRST2002 densities.
For the fragmentation functions we rely on the DSS \cite{DSS} set that provides 
full flavor and charge separation at NLO. Similar results for pions are obtained with the latest
AKK set \cite{AKK}. Considering that the process depends on two different hard scales, the transverse momentum
of the hadron and the one of the jet, we define the `default' scale to be the average of both of 
them, and, unless otherwise stated we set all the factorization and renormalization scales to
$\mu_F=\mu_R=\left( p_T^{\pi} +p_T^{jet}\right)/2$.

Fig. \ref{k} shows the result for the unpolarized (solid) and polarized (dashes) `$K-$factors' computed at the default scales  
 in terms of $p_T^{\pi}/p_T^{jet}$, the ratio between the transverse momentum of the pion and
  the jet \cite{adam}. 
 The relevance of this dimensionless ratio will become clear in the next section. 
As can be observed, the NLO corrections are sizable in the unpolarized case, ranging from 
50\% to more than 100\% of the Born result. The corrections are smaller in the polarized case,
as usual, resulting in an important decrease in the corresponding spin asymmetry. In that sense the 
situation is quite similar to the one found for single-hadron production \cite{singlehad,Jager:2002xm}.
\begin{figure}[htb]
\vspace{0.5cm}
\begin{center}
\begin{tabular}{c}
\epsfxsize=12truecm
\epsffile{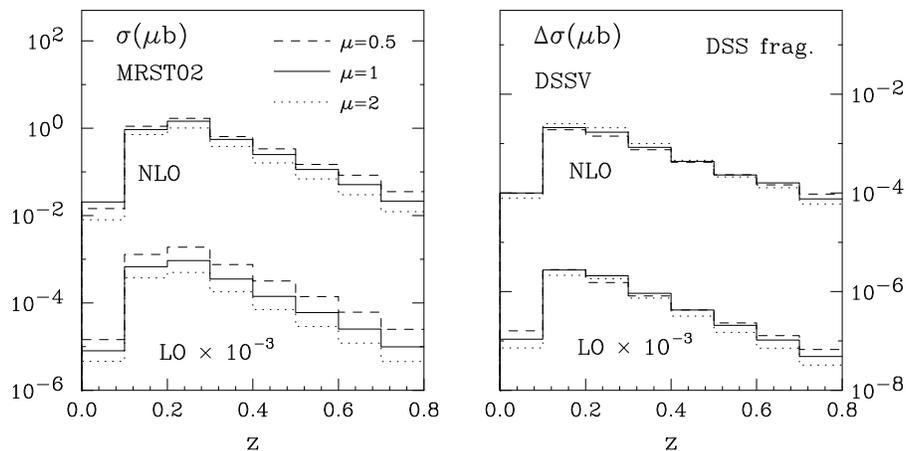}\\
\end{tabular}
\end{center}
\vspace{-0.5cm}
\caption{{\em \label{scales} LO and NLO unpolarized (left) and polarized (right) cross sections
at different factorization and renormalization scales $\mu_F=\mu_R=\mu \left( p_T^{\pi} +p_T^{jet}
\right)/2$. The LO results were scaled by a factor of $10^{-3}$ for better presentation.}}
\end{figure}
In some extreme kinematical regimes, the QCD corrections tend to be dominated by the 
contributions arising from soft-gluon emission that need to be resummed to all orders in the
coupling constant $\alpha_s$ to allow for a quantitative study.
It has been shown \cite{deFlorian:2005yj} that for inclusive single-hadron production  the  resummation
of the dominant terms is required in the case of
fixed-target energy experiments, where the `$K$-factors' largely exceed those found here, 
but not for a collider running at $\sqrt{s}=200$ GeV. 
Particularly, the effect of the resummation over the corresponding asymmetries is rather small 
\cite{de Florian:2007ty}. Considering that the dominant soft contributions for the 
hadron+jet observable originate from the same Sudakov form factors as in the inclusive case,
we believe those effects can also be neglected in a  first approach here.

A reliable error estimate on our NLO results requires some knowledge on the
size of the uncalculated higher-order terms. 
 The best we can do, before higher-order terms are computed,
  is to study  the dependence of the full NLO  results
on the renormalization and factorization scales.
Although physical
observables are obviously independent of the scales, theoretical predictions do
have such a dependence, arising from the truncation of the perturbative
expansion at a fixed order in the coupling constant $\alpha_s$. A large
dependence on the scales, therefore, implies a large theoretical uncertainty. 
In order to show how the scale dependence is substantially reduced once
the next-to-leading order corrections are included we will compare to the Born result.
For the sake of presentation we set all the scales to be equal, and vary them by a factor
of 2 up and down with respect to the default choice, i.e, $\mu_F=\mu_R=\mu \left( p_T^{\pi} +p_T^{jet}
\right)/2$ with $\mu=1/2,1,2$.
Fig. \ref{scales} plots the corresponding scale dependence of the unpolarized (left) and 
polarized (right) cross sections, were we observe a considerable reduction when the NLO 
corrections are included. Nevertheless, it is worth
 noticing that the scale dependence is still rather large at NLO for the unpolarized cross section,
 of the order of $\pm 20\%$ or more (compared to about $\pm 80\%$ at the Born level).
 The scale dependence is much smaller in 
 the polarized case, even reaching the stage in which at some kinematics the NLO cross section 
 evaluated at $\mu=2$ is larger than the one at $\mu=1/2$, opposite to the LO expectation.
 Since the 
 uncertainty in the unpolarized cross section directly contributes to the one for the asymmetry, 
 one might consider the convenience of using directly $\Delta \sigma$, instead of the asymmetry, 
 to extract the polarized parton distributions with a considerably better theoretical accuracy.

\section{Phenomenology}  

\begin{figure}[htb]
\vspace{0.1cm}
\begin{center}
\begin{tabular}{c}
\epsfxsize=6truecm
\epsffile{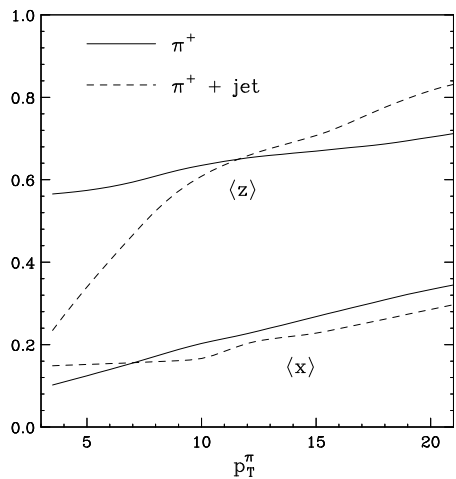}\\
\end{tabular}
\end{center}
\vspace{-0.5cm}
\caption{{\em \label{average} Average of the partonic momentum fractions $x$
 and $z$ 
for single pion production (solid) and pion accompanied by a jet with $p_T^{jet}>10$ GeV (dashes). }}
\end{figure}

As discussed in the Introduction, counting with the jet kinematics allows one
to impose cuts that enhance the relevance of some kinematical region in the momentum 
fractions $x_{1,2}$ and $z$. That can be observed in Fig. \ref{average}, where we 
plot the average 
value of the momentum $\langle x \rangle$ 
\footnote{The results are symmetric in the $x_1$ and $x_2$ distributions,
we denote it generally by $x$}
and hadronization $\langle z \rangle$ fractions for unpolarized single-$\pi^+$ 
production (solid) and the corresponding one for $\pi^+$+jet (dashes), in both cases in terms of 
the transverse momentum of the pion.
While the averages are relatively similar for $x$, the change is
quite clear for the fragmentation fraction $z$. Whereas for single-hadron production 
$\langle z \rangle\sim 0.6-0.7$  remains 
almost constant over the full kinematical range, it shows
a large variation, from $0.2$ to $0.8$, when the opposite jet is required. 
The rise of $\langle z \rangle$ with the transverse momentum of the pion can be easily understood from
simple physical considerations. At the Born level, only two final-state partons are produced,
with opposite transverse momentum. For that kinematics, the pion is produced by the fragmentation 
of one of the partons, while the jet is just formed by the other one. Therefore, the ratio between the 
transverse momentum of the pion and the one of the jet is exactly the hadronization 
fraction $z$. Once a jet cut is applied, selecting the transverse momentum of the pion is equivalent
to selecting the fraction of momentum that is transferred from the parton in the hadronization process. 
It is worth noticing that at the Born level, counting with the jet and hadron kinematics allows one to fully
reconstruct all the momentum fractions as
\begin{eqnarray}
\label{xz}
z &\equiv& \frac{p_T^{h}}{p_T^{jet}}  \nonumber  \\
x_1 &\equiv& \left(  p_T^{jet} \exp(\eta_{jet})+  p_T^{jet} \exp(\eta_{h})   \right)/\sqrt{s} \\
x_2 &\equiv& \left(  p_T^{jet} \exp(-\eta_{jet})+  p_T^{jet} \exp(-\eta_{h})   \right)/\sqrt{s} \nonumber .
\end{eqnarray}
\begin{figure}[htb]
\vspace{0.2cm}
\begin{center}
\begin{tabular}{c}
\epsfxsize=12truecm
\epsffile{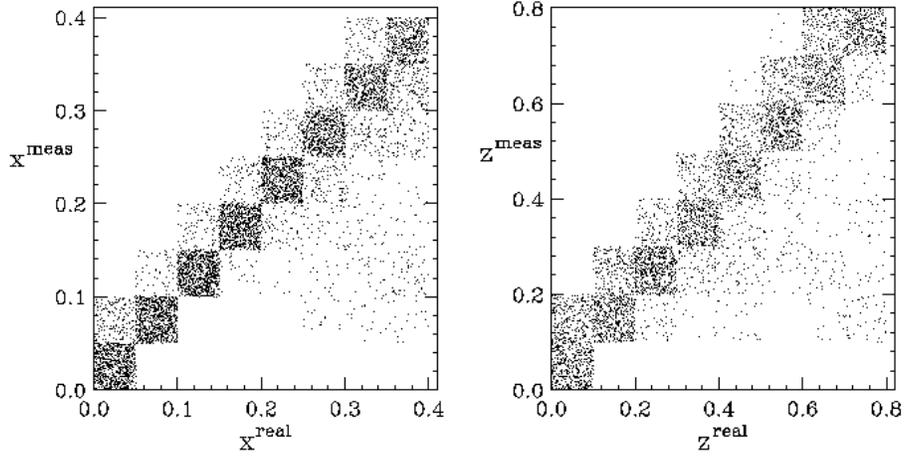}\\
\end{tabular}
\end{center}
\vspace{-0.5cm}
\caption{{\em \label{correlations} Correlations between the `measured' and `real' $x$ (left) 
and $z$ (right).}}
\end{figure}
While those relations are not valid at NLO, since one more parton can be radiated, still
one can observe that there is a strong correlation between the `real' momentum fractions
(the arguments of the parton distributions and fragmentation functions in Eq.(\ref{eq:fact})) and those 
obtained from the measured observables in Eq.(\ref{xz}). The correlations found for $\pi^+$ 
production in unpolarized collisions are plotted in Fig. \ref{correlations}. 
Considering a bin of size 
0.05 (0.1) for $x$ ($z$), we find that about 90\% (60\%) of the generated (weighted) events 
in the MonteCarlo implementation of the NLO corrections give
the same value for the `real' and `measured' momentum fractions, at least in the kinematical range where 
their contribution to the cross section is dominant.
\begin{figure}[htb]
\vspace{0.5cm}
\begin{center}
\begin{tabular}{c}
\epsfxsize=12truecm
\epsffile{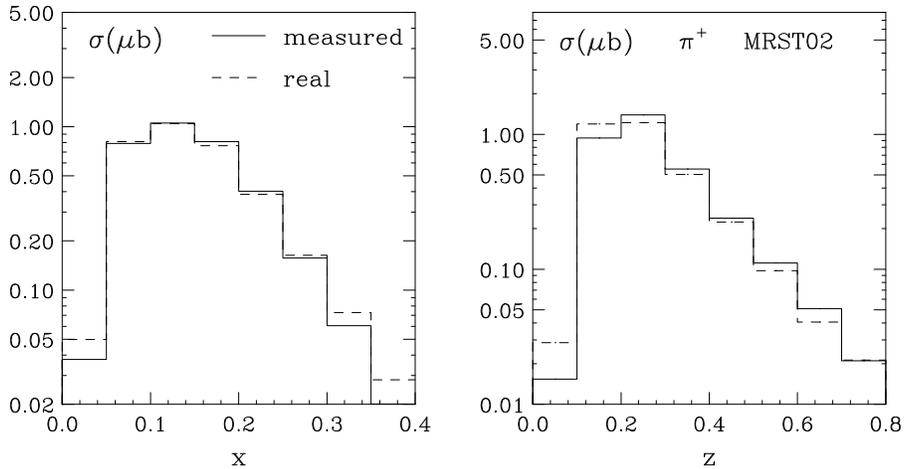}\\
\end{tabular}
\end{center}
\vspace{-0.5cm}
\caption{{\em  \label{xsection} NLO unpolarized cross section in terms of the `measured' (solid)
and `real' (dashes) partonic momentum fractions $x$ (left) and $z$ (right) }}
\end{figure}

The situation is also visible 
when the cross section is plotted in terms of the same variables, as shown in Fig. \ref{xsection}.
In the dominant range of $0.05\lesssim x \lesssim 0.3$, the agreement between the `measured' cross section and
the one obtained in terms of the `real' momentum fraction is at the percent level. 
The use of the variables in Eq.(\ref{xz}) can therefore allow for an accurate  
reconstruction of
the initial state momentum fractions $x_{1,2}$.
\begin{figure}[htb]
\vspace{0.5cm}
\begin{center}
\begin{tabular}{c}
\epsfxsize=12truecm
\epsffile{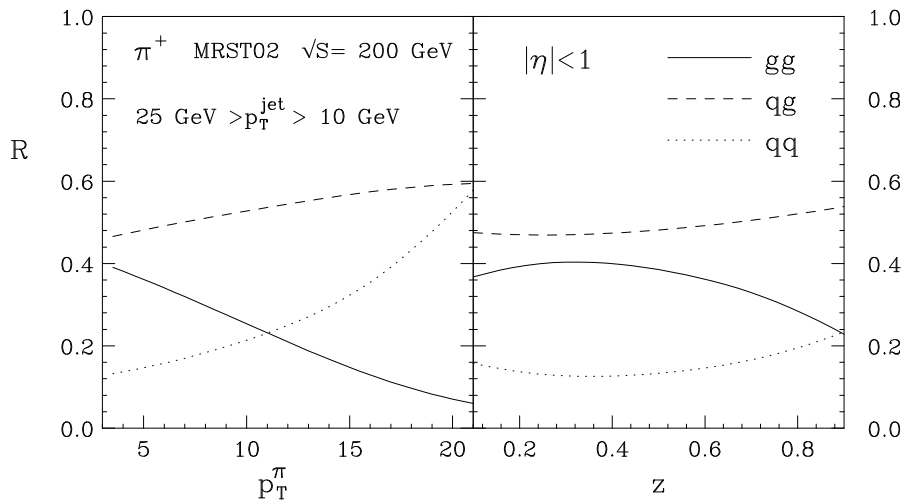}\\
\end{tabular}
\end{center}
\vspace{-0.5cm}
\caption{{\em \label{rates} Contribution to the cross section from the $gg$ (solid), 
$qg$ (dashes) and $qq$ (dots) 
channels in terms of the transverse momentum of the pion (left) and the variable 
$z=\frac{p_T^{\pi}}{p_T^{jet}}$ (right). }}
\end{figure}
In the case of 
the $z$ distributions, the differences between the `real' and the `measured' quantities 
can reach up to $10\%-15\%$. 
\begin{figure}[htb]
\vspace{0.5cm}
\begin{center}
\begin{tabular}{c}
\epsfxsize=7truecm
\epsffile{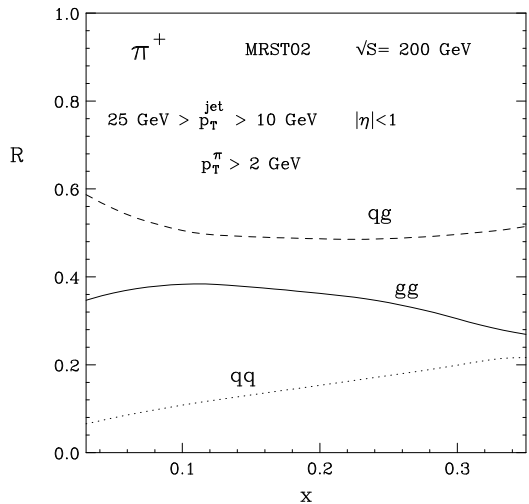}\\
\end{tabular}
\end{center}
\vspace{-0.5cm}
\caption{{\em  \label{ratesx}  Contribution to the cross section from the $gg$ (solid), 
$qg$ (dashes) and $qq$ (dots) 
channels in terms of $x$ }}
\end{figure}
However, it still becomes quite useful to plot the 
cross section in terms of it. This is mainly because, by selecting a range in $z$, one can enhance 
or decrease the contribution from some partonic channel due to the particular behavior of the
fragmentation functions. This feature can be observed in Fig. \ref{rates}, where we show 
the fractional contribution to the NLO unpolarized cross section from the $gg$, $qg$ and $qq$
initial state partonic channels \footnote{We include all different quark-(anti)quark channels under 
the common label $qq$.}. If the cross section is analyzed in terms of the transverse momentum of
the hadron, as it happens for the single-inclusive case, the pure gluonic channel contribution
$gg$ becomes only sizable at small values of $p_T^{\pi}$ and then decreases rapidly, making the 
cross section  less sensitive on the gluonic content of the proton. The situation changes when the 
same results are studied in terms of the variable $z$. Here the $gg$ channel shows a larger 
fractional contribution over the entire kinematical range at expenses of a suppression of the pure quark 
channels, that at most account for only 20\% of the cross section, providing an ideal scenario to
extract $\Delta g$ in polarized collisions.
Something similar occurs for the same observable plotted in terms of the variable $x$, as 
shown in Fig. \ref{ratesx}. The analyses confirm that hadron+jet production in hadronic
collisions, in terms of both dimensionless variables $x$ and $z$, provides a clear  source 
of information on the gluon distribution. In the next section we will look directly at the
correspondent sensitivity on $\Delta g$ in polarized $pp$ collisions.

\section{Asymmetries at RHIC and sensitivity on $\Delta g$}  

\begin{figure}[htb]
\vspace{0.5cm}
\begin{center}
\begin{tabular}{c}
\epsfxsize=12truecm
\epsffile{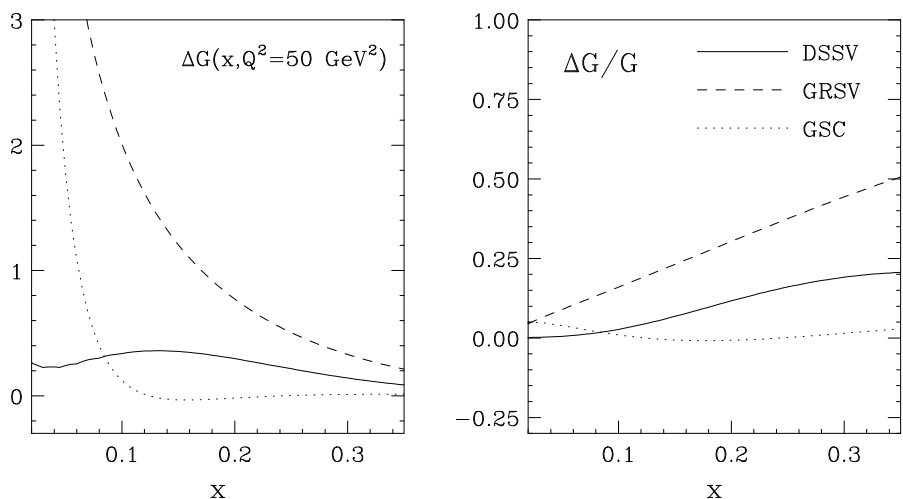}\\
\end{tabular}
\end{center}
\vspace{-0.5cm}
\caption{{\em  \label{pdf} Polarized gluon density at $Q^2=50$ GeV$^2$ from different sets of polarized pdfs 
(left) and their ratios to the unpolarized distribution (right).}}
\end{figure}

In order to analyze the sensitivity of the process on the polarized gluon distribution, we will 
compute the NLO asymmetries with three different sets of spin-dependent densities: DSSV \cite{dssv},
GRSV (standard) \cite{grsv}, and
GS-C \cite{gs}. The corresponding NLO distributions at $Q^2=50$ GeV$^2$, a typical scale for
this process, are shown in the left-hand side of Fig. \ref{pdf}. As can be observed,
the expectations from the three sets are quite different.
While the DSSV distribution corresponds to the best fit from the latest global analysis of all
polarized data \cite{dssv}, the GRSV set can be considered as an `upper bound' for the 
allowed range of gluon densities. The GS-C set provides a distribution  compatible with
the requirement of a small gluon polarization in the range $0.05\lesssim x\lesssim 0.3$ but with a node in that region
and a very different behavior at smaller $x$ compared to the DSSV set.
The right-hand side in Fig. \ref{pdf} shows the ratio between the corresponding 
polarized distribution and the unpolarized MRST2002 set.
In the quark sector, the dominant distributions are very
similar among the different sets. Therefore, since the
variations in the quark sector are much smaller than the ones for gluons, 
we can expect that any differences between predictions for the  asymmetries
that are found when using different 
polarized parton density sets are to be attributed to the 
sensitivity of the observable to $\Delta g$. 
\begin{figure}[htb]
\vspace{0.5cm}
\begin{center}
\begin{tabular}{c}
\epsfxsize=14truecm
\epsffile{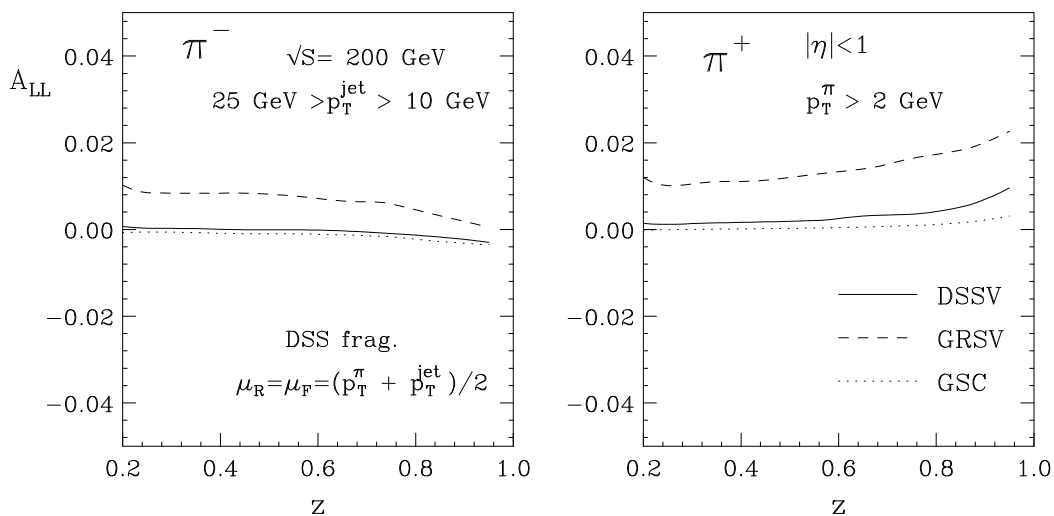}\\
\end{tabular}
\end{center}
\vspace{-0.5cm}
\caption{{\em \label{asymthz} Expected asymmetries for $\pi^-$ (left) and $\pi^+$ (right) 
production at RHIC from different sets of polarized pdfs in terms of $z$.}}
\end{figure}

We start the presentation of the expected asymmetries by looking first at $\pi^-$ and $\pi^+$ 
production in terms of the variable $z$,  as shown in Fig. \ref{asymthz}.
As expected, the asymmetries for the DSSV and GS-C distributions turn out to be small. 
When a set with a larger gluon distribution, like GRSV,
is considered, the asymmetries increase to the 1\%-2\% level.
The asymmetries for positive pions show a stronger sensitivity on the
polarized gluon distribution at large $z$, in line with similar findings for single pion production at 
large transverse momentum \cite{singlehad}. The differences and similitudes between negative and 
positive pion asymmetries can be easily understood: at small $z$, the 
`favored' (like $u\rightarrow \pi^+$ and $d\rightarrow \pi^-$)  and
`unfavored' (as $\bar{u}\rightarrow \pi^+$ and $\bar{d}\rightarrow \pi^-$) fragmentation 
functions are rather similar and therefore the cross section is almost charge invariant in that range.
On the contrary, at larger $z$, favored distributions overcome
the unfavored ones. The asymmetries
reflect the differences between the positive $\Delta u$ and negative $\Delta d$ parton 
distributions in the polarized proton that contribute with a different weight to the cross section. 
In this regime, the $\Delta u \Delta g$ channel becomes dominant for $\pi^+$ production resulting
in a larger sensitivity on the polarized gluon density.
\begin{figure}[htb]
\vspace{0.5cm}
\begin{center}
\begin{tabular}{c}
\epsfxsize=14truecm
\epsffile{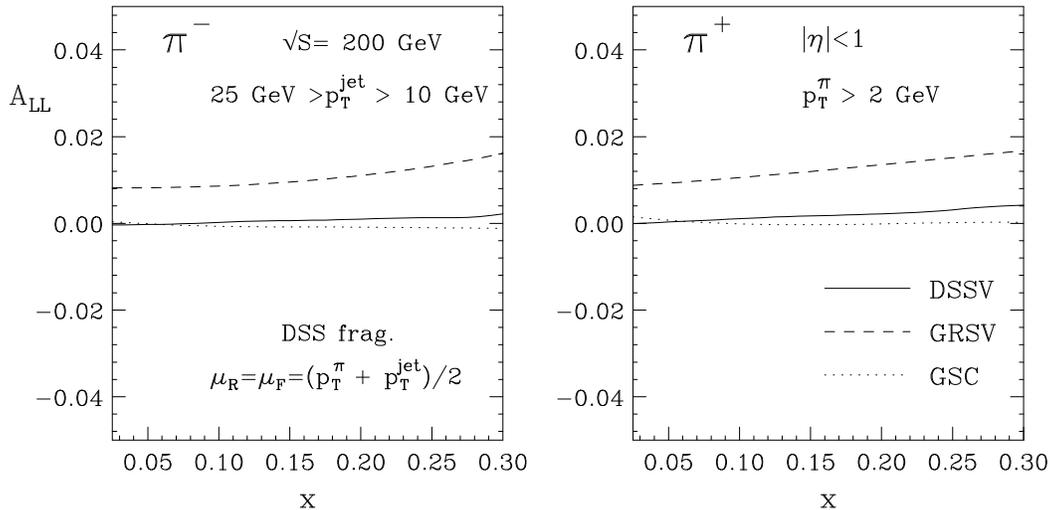}\\
\end{tabular}
\end{center}
\vspace{-0.5cm}
\caption{{\em \label{asymthx} Expected asymmetries for $\pi^-$ (left) and $\pi^+$ (right) 
production at RHIC from different sets of polarized pdfs in terms of $x$.}}
\end{figure}
A similar analysis can be performed in terms of the variable $x$, as defined in Eq.( \ref{xz}). Here, since
$z$ is integrated out, one can expect closer results for $\pi^-$ and $\pi^+$ production. 
This is observed in Fig. \ref{asymthx}, where both asymmetries reflect the shape and order of the 
curves for the 
$\Delta G/G$ ratio plotted in the right-hand side of Fig. \ref{pdf}. Notice that the sensitivity on
the gluon distribution for $\pi^-$ production is increased when the asymmetry is analyzed 
in terms of the variable $x$ instead of $z$.
\begin{figure}[htb]
\vspace{0.5cm}
\begin{center}
\begin{tabular}{c}
\epsfxsize=14truecm
\epsffile{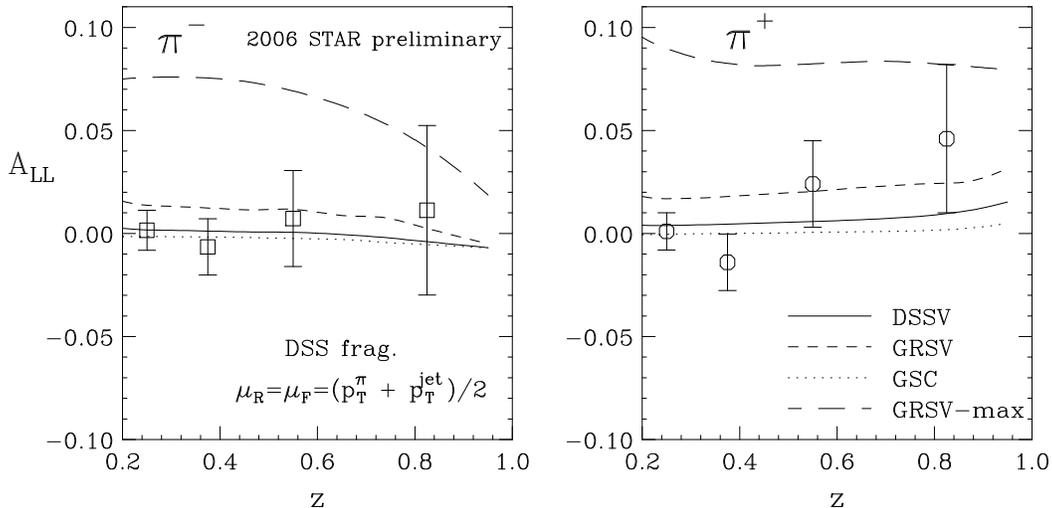}\\
\end{tabular}
\end{center}
\vspace{-0.5cm}
\caption{{\em \label{asymexp} Asymmetries measured by STAR at RHIC for $\pi^-$ (left) and $\pi^+$ (right) 
compared to the prediction from different sets of polarized pdfs. The theoretical predictions 
were corrected to account for the jet trigger efficiency. }}
\end{figure}

The STAR collaboration at RHIC has recently presented preliminary data on $\pi+$jet production.
Even though the data has been obtained with very similar cuts to those used along this work, the
experimental results can not be directly compared to the expected NLO asymmetries shown above  
because the data has not been corrected by the jet trigger efficiency. In order to allow for a comparison,
we have recomputed the corresponding asymmetries by incorporating in the theoretical calculation
the trigger efficiency parametrized as in Ref. \cite{adam} and modifying the experimental cuts 
accordingly.
The only modification with respect to the previous cuts are those applying to the jet,
which is required to have a transverse momentum in 9.5 GeV $<p_T^{jet}<$ 25 GeV   
and rapidity in the range $-0.7<\eta<0.9$ \footnote{The asymmetries are almost insensitive to the
precise values of those cuts.}.
The result is presented in Fig. \ref{asymexp}. The main effect of the trigger is to enhance the 
contribution from large $p_T^{jet}$ with respect to the small $p_T^{jet}$ events and, therefore, 
increase the average $\langle x \rangle$ resulting in larger asymmetries.

With the present experimental accuracy it is not yet possible to perform a precise extraction of the 
polarized gluon density from this observable. Nevertheless, the data can already rule out any 
possible scenario with a large gluon polarization in the range
$0.05\lesssim x\lesssim 0.3$. For that purpose we include in Fig. \ref{asymexp} the prediction from the set
GRSV-max set, where the polarized gluon distribution
is assumed to be equal to the unpolarized density at the very low intial scale 
of $\mu^2 =0.4$ GeV$^2$. 
That set completely overestimates the experimental data at small $z$.
Therefore, in line with other measurements performed
at RHIC, the preliminary data confirms the results from the global analysis in \cite{dssv} and 
points out to a small gluon polarization in the proton.

\section{Conclusions}

It is shown that the perturbative stability of the hadron+jet cross section 
improves considerably after including the NLO contributions. The corrections are found 
to be nontrivial: $K-$factors are larger for the unpolarized cross section 
than for the polarized one, resulting in a reduction of the asymmetry at NLO.
The possibility of looking at charged pions accompanied by a back-to-back jet 
is studied phenomenologically in detail, 
finding that the asymmetries for $\pi$ prodution,
in terms of both dimensionless variables $x$ and $z$,
are sensitive to the polarized gluon density in the range $0.05\lesssim x\lesssim 0.3$.
Furthermore, we show to NLO accuracy that the available data collected by the STAR 
collaboration at RHIC for this observable confirms the idea of a rather small gluon polarization.


  \subsection*{Acknowledgments}

This work was supported by UBACYT, CONICET, ANPCyT and the Guggenheim Foundation.
The author is grateful to Werner Vogelsang, Bernd Surrow and Adam Kocoloski for 
comments and discussions.
  

\end{document}